\documentclass[onecolumn]{revtex4}    % 正文双栏

\usepackage{graphicx}
\usepackage{mathrsfs}
\usepackage{amsmath}
\usepackage{CJK}
\usepackage{amsfonts} % 数学字体、符号,mathrsfs,amssymb
\begin{document}

\title{Logarithmic distribution of mean velocity and turbulent kinetic energy in a pipe flow}

%\title[A Lie-group derivation of mixing length formula] {A Lie-group derivation of a multi-layer mixing length formula for turbulent channel and pipe flow}

\author{Xi Chen$^{\rm a}$\vspace{6pt}, Fazle Hussain$^{\rm
a,b}$ and Zhen-Su She$^{\rm a}$$^{\ast}$\thanks{$^\ast$Corresponding
author. Email: she@pku.edu.cn \vspace{6pt}} \\\vspace{6pt} $^{\rm
a}${\em{ State Key Laboratory for Turbulence and Complex Systems and
Department of Mechanics and Aerospace Engineering, College of
Engineering, Peking University, Beijing 100871, P.R. China}}; \break
$^{\rm b}${\em{Department of Mechanical Engineering, University of
Houston, Houston, TX 77204-4006, USA}}\\\vspace{6pt}\received{May
2012} }

%\date{\today}

\maketitle

%\begin{abstract}
 A Lie-group based similarity theory is developed for both
 momentum and energy distributions in a turbulent pipe flow,
 leading to asymptotic logarithmic profiles of mean velocity
 and turbulent kinetic energy. Both channel and pipe data
 over a wide range of $Re$ yield $0.45$ to be the universal
 Karman constant. A new spatial invariant characterizing
 outer dynamics is discovered and validated by reliable
 experimental data. The theory predicts the mean velocity
 profile (MVP) with $99\%$ accuracy for high $Re$ experimental
 data (up to $40$ millions), and offers a quantitative
 explanation for recent observation of logarithmic kinetic
 energy distribution by Hullmak et al. (Phys. Rev. Lett. 108, 094501).
%\end{abstract}

%\pacs{47.27.-i, 47.27.N-, 47.27.E-} \maketitle

%\section{\label{sec:level1}First-level heading}
% sections are not used for PRL papers

Turbulent flows over objects form thin vorticity layers called
boundary layers. As it is widely accepted that near-wall flow
physics is autonomous and independent of the flow being external or
internal, pipe flow forms an experimentally and numerically
expedient canonical flow for the study of wall turbulence. Despite
extensive efforts, the prediction of the mean velocity still relies
on empirical functions \cite{Wilcox06} having limited accuracy and
limited range of Reynolds numbers ($Re$). Hence, the problem
continues to receive vivid attention with great experimental
\cite{Marusic2010, Smits2011} and theoretical \cite{Lvov2008}
efforts.

From a statistical physics point of view, turbulent pipe flows are
at a far-from-equilibrium state encompassing not only a cross-scale
energy flux (cascade) but also momentum and energy fluxes in space.
Understanding physical principles governing the non-homogeneous
transport and non-uniform distribution of the mean momentum and
kinetic-energy is a log-standing goal of the research. Nearly eighty
years ago, Prandtl \cite{Prandtl1925} and von Karman
\cite{Karman1930}, independently proposed the concept of mixing
length with a linear dependence on the distance from the wall,
predicting a logarithmic MVP and hence friction coefficient.
However, this empirical model has led to controversies: Barenblatt
et al \cite{Barenblatt} have claimed that power-law is a better
description; Goldenfeld \cite{Goldenfeld} has proposed a model for
friction coefficient using a power-law description. A recent model
of L'vov et al. \cite{Lvov2008} is particularly noteworthy, as its
log-law description yields predictions of reasonable accuracy over a
range of finite $Re$ (see Fig.2). Recently, a logarithmic scaling
for the streamwise mean kinetic energy profile (MKP) is reported
\cite{Hulkmark2012}, with no explanation. Clearly, a deductive
theory for joint MVP and MKP is still missing.

Here, we present a Lie-group based similarity theory for turbulent
channel and pipe flows. The original idea was presented in
\cite{She2010} and formulated rigorously in \cite{She2012}. The goal
of the theory is to find invariant solutions of the averaged flow
equations based on a symmetry analysis of a set of new quantities,
called \emph{order functions}, which are introduced in close analogy
to \emph{order parameter} \cite{Kadanoff2009} in the study of
critical phenomena. Adding the order function to dependent variables
in the equations and then performing a dilation-group transformation
yields a set of new, candidate invariant solutions, which defines a
new method to measure the Karman constant. In this Letter, we
compare the prediction to measured MVP in a turbulent pipe
\cite{Mekeon2004}. In addition, with a system similarity argument,
we find a new spatial invariant which predicts a logarithmic MKP at
high $Re$.

\emph{Theory for mean velocity} - In a pipe, the mean momentum
equation (MME) is \cite{Lvov2008}:
\begin{equation}\label{MME}
S^ + + W^ +=\frac{{dU^+  }}{{dy^ +  }} - \left\langle {uv}
\right\rangle ^ + = \tau _U^ +
\end{equation}
where  $U^+\equiv \left\langle u \right\rangle^+$ is the streamwise
mean velocity, $ S^ + = dU^ + /dy^ +$ is the viscous stress, $ W^+ =
- \left\langle {\left. {uv} \right\rangle } \right.^ +$ is the
Reynolds stress, and $\tau_U^+\equiv r$ is the total wall shear
stress, $r$ is the distance to the centreline, $y^+$ the distance to
the wall, and $u$, $v$ are streamwise and vertical fluctuating
velocities, respectively, and superscript $+$ denotes 'wall units'
normalization with friction velocity and viscosity. We now search
for group-invariant solutions of (\ref{MME}), by introducing the
mixing length, $\ell^+_M=\sqrt{W^+}/S^+$. A formal Lie-group
analysis of (\ref{MME}) adding $\ell_M$ ($=\ell^+_M/Re_\tau$) and
its gradient, $d\ell_M/dr$, as new dependent variables
\cite{She2012} shows three kinds of invariant solutions, of which
the second kind corresponds to the situation of symmetry-breaking in
$\ell_M$ but maintained symmetry in $d\ell_M/dr$. This solution is
expressed as:
\begin{equation}\label{lm-bulk}
\ell_M(r)\approx\frac{\kappa}{m}(1-r^m)
\end{equation}
where $m$ is a scaling exponent characterizing the bulk flow, and
$\kappa$ is the classical Karman constant, as can be seen when
taking the limit to the wall, $r\rightarrow1$, $\ell_M \approx
\kappa y$, i.e. Karman's linearity assumption.

Now, we establish a complete expression for the mixing length in the
outer flow encompassing a bulk and a central core. Define another
length involving $S^+$, $W^+$ and dissipation $\epsilon^+$, using
the eddy viscosity $\nu_t=W^+/S^+$ \cite{Prandtl1925}, which gives $
\ell _\nu ^ + = (W^+/S^+)^{3/4}/\varepsilon^{+1/4} = \nu _t^{3/4}
/\varepsilon ^{ + 1/4}$. A comparison to the Kolmogorov dissipation
length $\eta=\nu^{3/4}/\varepsilon^{1/4}$ suggests that $\ell_\nu$
signifies the energy input length scale, while $\eta$ the output
length scale of the energy cascade. Verify that
\begin{equation}\label{lm-lnu}
\ell _M^ + = \ell _\nu ^ + \Theta ^{1/4},
\end{equation}
where $ \Theta  = \varepsilon ^ + /(S^ + W^ +  ) = \varepsilon ^ +
/P^ +$ is the ratio between dissipation and production. $\Theta$ is
an important quantity, denoted to be an order function of the second
kind \cite{She2010}.

A two-layer model can be readily derived with the above definitions.
In the bulk flow, the quasi-balance \cite{Pope00} corresponds to $
\Theta\approx 1$. Near the center line as $r\to 0$, $W^+\approx r$,
$S^+\approx \sqrt{r}/\ell_M^+\sim r$ and $\varepsilon>0$, then
$\ell_M\sim r^{-1/2}$, but $\ell_\nu\to C=O(1)$ and $\Theta\sim
r^{-2}$. This transition is physically due to the switch of the
generating mechanism of fluctuations - from mean shear production to
turbulent transport (vanishing mean shear at the center). A simple
ansatz taking into account of the transition of $\Theta$ from
$r^{-2}$ to $1$ is $\Theta = 1 - c + cr^{-2}$ with a constant $c$,
or
\begin{equation}\label{Theta}
\Theta = \frac{1+(r_c/r)^2}{1+r_c^2},
\end{equation}
where the parameter $ r_c  = \sqrt {c/(1 - c)}$ represents a
critical radius characterizing the core region. In this core region
($r_c>r\to 0$), $\ell_\nu\approx \ell_0$ and $\Theta\approx
r^{-2}r_c^2/(1+r_c^2)$ together determine the central behavior of
$\ell_M$ from (\ref{lm-lnu}).

In the bulk flow ($r>r_c$), $\Theta\approx 1$,
$\ell_\nu\approx\ell_M$. A phenomenology yields an estimation of $m$
in (\ref{lm-bulk}) as following. Consider the scaling property of
$\ell _\nu$. Similar to $\ell_M$, only $d\ell_\nu/dr$ has a
scale-invariance property: $d\ell_\nu/dr\propto r^{m-1}$, which
corresponds to the Lie-group similarity of the second kind
\cite{She2012}. The exponent $m$ can be derived by assuming that: $
d\ell_\nu/dr \propto \int {P^+} rdr$, i.e. the volume integral of
the turbulence production, which equals the amount of kinetic energy
converted from the mean flow. The validity of this assumption relies
on an intriguing connection between the mean flow and fluctuation
energy yet to be uncovered, which we defer to future study. Since
$P^+=S^+W^+\propto r^2$ and using the wall condition ($\ell_\nu=0$
as $r\to 1$), we obtain $\ell _\nu = \ell _0(1-r^5)$. Furthermore,
taking the limit $ r \to 1$ and using $\ell_M\approx \kappa y$ and
$\Theta\approx 1$, we obtain an important relation: $\kappa  = 5\ell
_0$. Hence, $ \ell _\nu   = \kappa (1 - r^5 )/5$. This derivation
immediately predicts that, for channel flow, $m=4$, since
$d\ell_\nu/dr \propto \int {P^+} dr$ with a flat plate. Substitute
the expression of $\ell_\nu$ and (\ref{Theta}) into (\ref{lm-lnu}),
we thus obtain:
\begin{equation}\label{lmout}
\ell _M (r) = \frac{{\kappa }}{{mZ_c }}(1 - r^m )\left( {1 + ({{r_c
}}/{r})^2 } \right)^{1/4}
\end{equation}
where $Z_c= (1 + r_c^2 )^{1/4}$, and $m=5$ (4) for pipe (channel)
flows.

\begin{figure}
\begin{center}
\includegraphics[width=9cm]{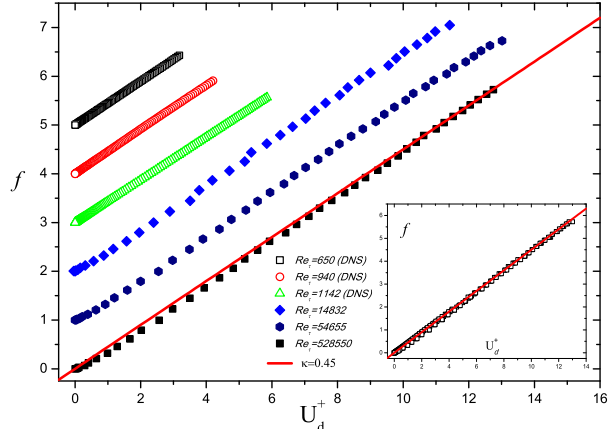}
\caption{(color). Plot of measured $U^{+}_d$ versus theoretical
$f(r; r_c)$ (vertically shift by one) for five sets of data : DNS
channel (black and red; $r_c=0.27$), DNS pipe (green; $r_c=0.23$),
experiments pipe (solid diamonds, disks and squares; $r_c=0.67$).
Note a good linear relation with slope $0.45$ for all $\emph{Re}$s.
The original plot is shown in the inset.} \label{fig:kappa}
\end{center}
\end{figure}

\emph{Measurement of $\kappa$ and prediction of the MVP} - In the
outer flow, $W^+\approx \ell_M^{+2}S^{+2}\approx r$, which yields
$S^ + \approx \sqrt r /\ell _M^ +$. Integrating it using
(\ref{lmout}) yields an expression for the mean velocity defect,
i.e. $ U_d^+ = U_c^+ - U^+(r) ={\mathop{\rm Re}\nolimits}_\tau
\int_0^r {S^ + dr} =(1/\kappa)f(r; r_c)$, where $ U_c^ +$ is the
mean velocity at the centreline, and
\begin{equation}\label{lm-f}
f(r; r_c)=mZ_c \int_0^r {\frac{{r'dr'}}{{(1 - r'^m )(r'^2 + r_c^2
)^{1/4} }}}= \kappa U_d^+.
\end{equation}
The linear relation between $f(r; r_c)$ and $U_d^+$ can be subjected
to experimental tests, with a fixed $r_c$. In Fig.\ref{fig:kappa},
theoretical $f(r; r_c)$ versus empirical (measured) $U_d^+$ is
plotted for a wide range of data, from direct numerical simulations
(DNS) of channel and pipe flow at moderate $Re$ \cite{Iwamoto2002,
Hoyas2006, Moin} and experiments of pipe flows \cite{Mekeon2004} at
very high $Re$. The linearity is remarkably observed with a slope of
$0.45\pm0.01$, consistent with our earlier study of channel flows
\cite{WY2012b}. Note that $r_c$ can be determined in a rational way,
by an evaluation of the minimum relative error, $\sigma _U  =
\frac{1}{N}\sum {\left( {1 - \kappa U_d^{+EXP} (r_i )/f(r_i; r_c)}
\right)^2 }$, where $\kappa$ is the value from the least-square
fitting above. In practice, the relative error is evaluated in a
domain defined as $200 < y^ + < 0.9{\mathop{\rm Re}\nolimits} _\tau
$. This procedure yields a $r_c\approx 2/3$ for Princeton data
\cite{Mekeon2004} at high $Re_\tau>5000$, while at the moderate to
low Re, the DNS data \cite{Iwamoto2002, Hoyas2006, Moin} show a
smaller $r_c\approx 0.23\sim 0.27$ for $Re_\tau\le 1000$ (both
channel and pipe). With $r_c$ so determined, $\kappa$ (i.e. the
slope) is measured with high confidence: all data show that
$\kappa\approx 0.45$.

This value of $\kappa$ is 10\% higher than generally accepted value
($0.41$) \cite{Wilcox06}. Initially, it was a bit surprising, but a
careful scrutiny confirms the internal consistency of the procedure.
Note that in our definition, $\kappa$ is a coefficient defining the
outer flow, and its measurement involves little ambiguity. The fact
that the measured value is accurately constant for both channel and
pipe flows and for a wide range of $Re's$, and that it is exactly
the historic Karman constant in the overlap region, suggests that
(\ref{lmout}) is a better definition for $\kappa$ in channel and
pipe flows, and taking into account explicitly the form of the outer
flow makes the measurement of $\kappa$ more robust, compared to
previous measurements \cite{Marusic2010, Nagib}. Another interesting
prediction is an asymptotic centreline dissipation at high $Re$:
\begin{equation}
\varepsilon _0^{ + Pipe(CH)}  = \mathop {\lim }\limits_{r \to 0} (S^
+  W^ +  \Theta ) = \frac{{mr_c^{3/2} }}{{\kappa Z_c^3 {\mathop{\rm
Re}\nolimits} _\tau  }} \approx \frac{4.6(3.7)}{{\mathop{\rm
Re}\nolimits} _\tau}.
\end{equation}
The constant 3.7 for channel flow becomes 1.18 at $Re_\tau\approx
940$ (with $r_c\approx 0.27$), which agrees with measured value
$\varepsilon_0^{+DNS}\approx 1.05/Re_\tau$ from DNS data of
\cite{Hoyas2006}.

\begin{figure}
\begin{center}
\includegraphics[width=9cm]{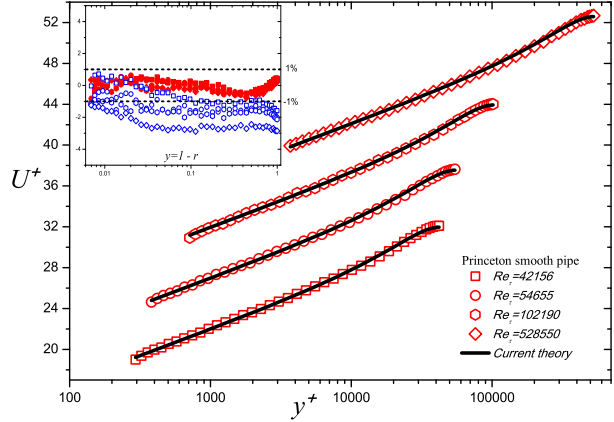}
\caption{(color). Theoretical (solid lines) and measured MVPs, which
are staggered vertically by five units for clarity. Inset shows the
relative errors, $ (U^{EXP} /U^{THE}-1) \times 100\;\%$  - our
predictions (red solid symbols) are uniformly within 1\%. Also
included (blue open symbols) is the model of L'vov et al.
\cite{Lvov2008}, which reveals systematic deviations at high $Re$
due to inappropriate boundary constraints \cite{WY2012a}.}
\label{fig:MVP}
\end{center}
\end{figure}

In order to predict the MVP, one needs an additional constant. Using
(\ref{lm-f}), we can express the MVP as
\begin{equation}\label{U-sed}
U^+(r)=U^+_c-f(r; r_c)/\kappa \end{equation} Our analysis of the
Princeton pipe data at high $Re$ show that $U_c^+\approx \ln Re_\tau
/0.45+B_c$, with $B_c\approx 8.3$. With three parameters:
$\kappa\approx 0.45$, $r_c\approx 0.67$ and $B_c\approx 8.3$, the
high-$Re$ outer flow MVP is completely specified by (\ref{lm-f}) and
(\ref{U-sed}). In addition, an asymptotic calculation of
(\ref{lmout}) gives $ \mathop {\lim }\limits_{r \to 1} f(r;r_c ) =
\mathop {\lim }\limits_{r \to 1} \kappa U_d^ +  \left( r \right)
\approx  - \ln \left( {1 - r} \right) + \kappa B_d$ with
$B_d\approx1.7$, which only depends on $r_c$. Then, (\ref{lm-f})
yields an approximate log-law in the overlap region:
\begin{equation}U^+(r)\approx \ln
y^+/0.45 + 6.6
\end{equation}
where the additive constant is found from $B_c-B_d\approx 8.3-1.7$.
In Fig.\ref{fig:MVP}, the theoretical MVP (\ref{U-sed}) is shown to
agree with the Princeton pipe data for the entire profile with 99\%
accuracy for $Re$ up to $4\times10^7$ (better than \cite{Lvov2008}).
This unprecedented accuracy settles the debate between the
logarithmic law and power law with a rational description of the
bulk flow; it further shows that turbulence in pipe flows indeed
admits an analytic solution.

\emph{Predictions for fluctuations} - The mean kinetic-energy
equation (MKE) \cite{Pope00} can be rearranged in a form similar to
(\ref{MME}) as:
\begin{equation}\label{MKE}
 - \frac{{dK^+  }}{{dy^ +  }} + \left\langle {uuv} \right\rangle ^ +   = \tau _K^ +
\end{equation}
where $K^+\equiv \left\langle {uu} \right\rangle^+$ is the
streamwise kinetic energy, and $ \tau _K^ +   = \int_0^{y^ +  } {(S^
+  W^ + - \varepsilon ^ + _K)dy^{ + '} }  - \left\langle {pu}
\right\rangle ^ +$ involves the integration of turbulent production,
$SW$, dissipation, $ \varepsilon_K$, and a term due to pressure
transport, $\left\langle {pu} \right\rangle ^ + $. Note that one
often assumes a quasi-balance, then, $\tau_K^+$ would be small.
Detailed examination of empirical data shows that the integrated
deviation from the quasi-balance, although small, is fully
responsible for the non-uniform distribution of the kinetic energy.

\begin{figure}
\begin{center}
\includegraphics[width=9cm]{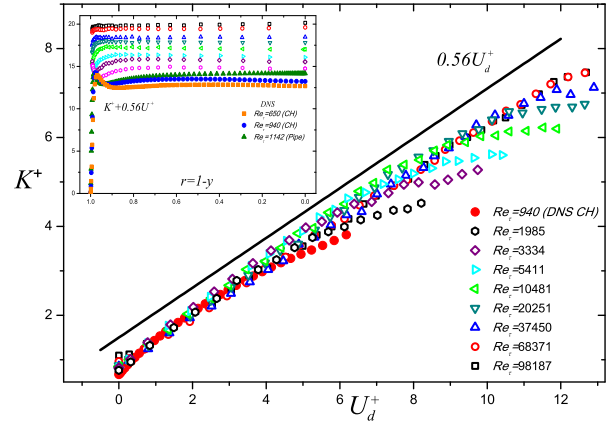}
\caption{(color). Evidence of linear dependence between $U^+_d$ and
$K^+$. The coefficient $\alpha\approx0.56$ is obtained from fitting
the largest $Re$ data. The inset validates the invariance law
(\ref{uub}) using experimental $K$ \cite{Hulkmark2012} and
theoretical $U$ at the same $Re$ with $\alpha$.} \label{fig:CST}
\end{center}
\end{figure}

\begin{figure}
\begin{center}
\includegraphics[width=9cm]{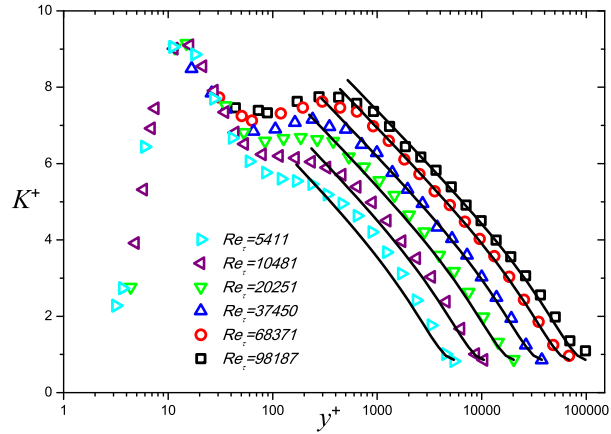}
\caption{(color). Comparison between the theory (\ref{MKP}) (solid
lines) and experimental data (symbols) \cite{Hulkmark2012} of the
MKP. } \label{fig:uu}
\end{center}
\end{figure}

The similarity between (\ref{MME}) and (\ref{MKE}) is a major focus
of this paper. Denote $S_K^ +   =  - dK^ +  /dy^ +$ and $ W_K^ + =
\left\langle {uuv} \right\rangle^+$, thus, the MKE has a similar
form as
\begin{equation}\label{MainEquation}
 S_K^+ + W_K^+= \tau_K^+.
\end{equation}
The two terms on the l.h.s. represent the viscous diffusion and
turbulent transport of $K$. Analysis of DNS data of channel flows
\cite{Hoyas2006} reveals indeed two similarities (results not
shown): first, all terms go to zero near the center line, and
$W^+\sim W^+_K$ and $S^+\sim S^+_K$; second, $W^+_K\gg S^+_K$, but
$W^+/S^+\sim W^+_K/S^+_K$. These similarities suggest that
(\ref{MME}) and (\ref{MKE}) may differ by a constant factor
$\alpha$. Assuming this is true, multiplying (\ref{MME}) by $\alpha$
and subtracting (\ref{MKE}) yields $d(\alpha U^++K^+)/dy^+\approx
0,$
%\begin{equation}\label{uua}
%\alpha U_d^ +  \left( r \right) + K_d^ +  \left( r \right) \approx 0,
%\end{equation}
leading to
\begin{equation}\label{uub}
\alpha U^ + \left( r \right) + K^ +  \left( r \right) \approx \alpha
U_c^ +   + K_c^ +\approx C
\end{equation}
where $K^+_c$ is central kinetic energy, with
\begin{equation}\label{C}
C=(\alpha/\kappa)\ln {Re_\tau}+\alpha B_c+K^+_c\end{equation} being
a constant in the outer flow. With experimentally measured $U^+(r)$
and $K^+(r)$, the validity of (\ref{uub}) is successfully tested
(Fig.\ref{fig:CST}) with a $\alpha\approx 0.56$ (Fig.\ref{fig:uu})
by a linear fitting at small $r$ (where the linearity is accurate
because both terms have a quadratic dependence on $r$).
Fig.\ref{fig:CST} shows a clear evidence of a spatial invariant over
an increasing radial domain with increasing $Re$: the extent reaches
almost the entire radius at high $Re$. In addition, the constant $C$
can be measured directly from data at each $Re$, shown in the inset:
$C\approx 1.25\ln Re_\tau +5.4$. This measurement yields
$K^+_c\approx0.7$.

A specific prediction of (\ref{uub}) is that at high $Re$, $K$ must
have a logarithmic profile (with a negative sign), since $U$ has a
logarithmic profile. In the overlap region, $K^ + \approx  - (\alpha
/\kappa )\ln y + \alpha B_d  + K_c^ + \approx -1.25 \ln y +1.65$,
which reproduces well the empirical observation of Hulkmark et al.
\cite{Hulkmark2012}: $ K_{}^ + \approx - 1.25\ln y  + 1.61$.
Finally, we predict the outer profile of the MKP as
\begin{equation}\label{MKP}
K^ + \approx \alpha f(r; r_c)/\kappa + K^+_c.
\end{equation}
As shown in Fig.\ref{fig:uu}, it agrees well with empirical data,
especially at high $Re$, better than that proposed by Alfredsson et
al \cite{Alfredsson2011}.

In summary, we have achieved a simultaneous description of the outer
MVP and MKP by exploring the similarity in the MME and MKE, which
discovers a new spatial invariant in the radius direction of a
turbulent pipe. The Lie-group based theory yields a new procedure
for measuring $\kappa$ and a universal value of $0.45$ for both
channel and pipe flows. The predicted MVPs achieve a 99\% accuracy
compared to Princeton pipe data for a wide range of $Re's$. Note
that the analysis has been successfully extended to incompressible,
compressible and rough-wall turbulent boundary layers
\cite{SED2012a, SED2012b}, and turbulent Rayleigh-Benard convection
(temperature). Those results will be communicated soon.

%\begin{figure}
%\begin{center}
%\includegraphics[width=10cm]{figure/CST}
%\caption{(color). Empirical evidence of the invariance law
%(\ref{uub}) by experimental and DNS data
%\cite{Hoyas2006,Iwamoto2002, Moin2008} of both channel and pipe
%flow. Note that the zone of the conservation extends to almost the
%whole pipe at high $Re$. In the inset, the plateau has a logarithmic
%dependence on $Re_\tau$. The data of $U_d$ and $U$ in this figure
%and Fig.\ref{fig:MKP} are from the dataset of McKeon et al
%\cite{Mekeon2004}, by choosing the closest $Re$ to that of $K$.}
%\label{fig:CST}
%\end{center}
%\end{figure}

We thank Y. Wu for helpful discussions. This work is supported by
National Nature Science Fund 90716008 and by MOST 973 project
2009CB724100.

$\ast$ Corresponding author. Email: she@pku.edu.cn


\begin{thebibliography}{99}


\bibitem {Wilcox06} D.C. Wilcox, 2006 {\it Turbulence Modeling for CFD.} DCW Industries.

\bibitem{Marusic2010} I. Marusic, \emph{et al.}, {Phys. Fluids.} \textbf{22}, 065103 (2010).

\bibitem{Smits2011}  A.J. Smits, B.J. McKeon, I. Marusic, {Annu. Rev. Fluid Mech.} \textbf{43}, 353-75 (2011).

\bibitem{Lvov2008} V.S. L'vov, \emph{et al.}, {Phys. Rev. Lett.} \textbf{100}, 050504 (2008).

\bibitem{Prandtl1925} L. Prandtl, {Z. Angew. Math. Mech.} \textbf{5}, 136-139 (1925).

\bibitem{Karman1930} von Karman, {\it In Proc. Third Int. Congr. Applied Mechanics}, Stockholm. 85-105.

\bibitem{Barenblatt} G. I. Barenblatt,  A. J. Chorin, {\it Proc. Natl Acad. Sci.} USA 101, 15023-15026 (2004).

\bibitem{Goldenfeld} N. Goldenfeld, {\it Phys. Rev. Lett.} 96, 044503 (2006).

\bibitem{Hulkmark2012} M. Hulkmark, et al., Phys. Rev. Lett. 108, 094501(2012).

\bibitem{She2010} Z.S. She, X. Chen, Y. Wu, F. Hussain, {Acta Mech Sinica}, 26, 847-861 (2010)

\bibitem{She2012} Z.S. She, X. Chen, F. Hussain, arXiv:1112.6312 (2012)

\bibitem{Kadanoff2009} L.P. Kadanoff, {\it J. Stat. Phys.} \textbf{137}, 777-797 (2009).

\bibitem{Mekeon2004} B.J. McKeon, et al., {\it J. Fluid Mech.} 501, 135 (2004)

\bibitem{Pope00} S.B. Pope, {\it Turbulent Flows.} (Cambridge University Press, Cambridge, 2000).

\bibitem{Iwamoto2002} K. Iwamoto, Y. Suzuki, N. Kasagi, THTLAB Internal Report. No. ILR-0201 (2002).

\bibitem{Hoyas2006} S. Hoyas, J. Jimenez, {\it Phys. Fluids.} \textbf{18}, 011702 (2006).

\bibitem{Moin} X.H. Wu, P. Moin, {\it J. Fluid Mech.} 608, 81-112 (2008).

\bibitem{WY2012b} Y. Wu, X. Chen, Z.S. She and F. Hussain, {\it Physica Scripta}, to appear (2012).

\bibitem{Nagib} H.M. Nagib, K.A. Chauhan, {\it Phys. Fluids.} 20, 101518 (2008).

\bibitem{WY2012a} Y. Wu, X. Chen, Z.S. She, F. Hussain, {\it Sci China-Phys Mech Astron}, 55, 9, 1691-1695 (2012).

\bibitem{Alfredsson2011} P.H. Alfredsson, \emph{et al.}, {Phys. Fluids.} \textbf{23}, 041702 (2011).

\bibitem{SED2012a} Y.S. Zhang, et al., {\it Phys. Rev. Lett.} 109, 054502 (2012).

\bibitem{SED2012b} Z.S. She, et al., {\it New J. Physics}, accepted, (2012).

%\bibitem{Ahlers09} G. Ahlers, S. Grossmann, D. Lohse, {\it Rev. Mod.
%Phys.} 81, 503-537 (2009).



%以下数字是用来标示空余的行数。只要在4页之内，留够24行，文章长度就满足要求了。





\end{thebibliography}
\end{document}